\PassOptionsToPackage{dvipsnames}{xcolor}
\documentclass[longbibliography,nofootinbib,aps,prx,superscriptaddress,twocolumn]{revtex4-2}

\usepackage{amsmath,amsfonts,amssymb,amsthm,dcolumn,bm,qcircuit,appendix,mathtools,thmtools,thm-restate,algorithm,algpseudocode,mathrsfs,pgfplots,soul,lipsum,graphicx,braket,enumitem,caption,subcaption,ragged2e}

\usepackage{fontawesome}

\usepackage{hyperref}

\usepackage[
  labelfont=bf,
  labelsep=period,
  justification=justified,
  format=plain,
  singlelinecheck=false
]{caption}

\usepackage{pgfplots}
\usepgfplotslibrary{patchplots}
\usetikzlibrary{3d, decorations.pathmorphing}
\usepackage{tikz,tkz-euclide,tikz-3dplot}
\pgfplotsset{compat=newest}
\usepgfplotslibrary{colormaps}
\usepgfplotslibrary{patchplots}

\usetikzlibrary{matrix,fit,backgrounds,3d,arrows, decorations.pathreplacing,shapes.geometric,3d, calc}

\def\BibTeX{{\rm B\kern-.05em{\sc i\kern-.025em b}\kern-.08em
    T\kern-.1667em\lower.7ex\hbox{E}\kern-.125emX}}

\newtheorem{theorem}{Theorem}
\newtheorem{lemma}{Lemma}

\newtheorem{remark}{Remark}
\newtheorem{definition}{Definition}
\newtheorem{method}{Algorithm}

\numberwithin{proposition}{section}
\numberwithin{theorem}{section}
\numberwithin{lemma}{section}
\numberwithin{corollary}{section}
\numberwithin{remark}{section}
\numberwithin{definition}{section}
\numberwithin{equation}{section}

\newcommand{\Ibb}{\mathbb{I}}

\newcommand{\Rbb}{\mathbb{R}}

\newcommand{\norm}[1]{\left\lVert#1\right\rVert}

\usepackage{xcolor}
\usepackage{hyperref}

\hypersetup{
    colorlinks=true,
    linkcolor=MidnightBlue,
    filecolor=magenta,
    urlcolor=teal,
    pdfpagemode=FullScreen,
    citecolor=OliveGreen
}

\begin{document}
\title{Quantum Kaczmarz Algorithm for Solving Linear Algebraic Equations}

\author{Nhat A. Nghiem}
\email{{nhatanh.nghiemvu@stonybrook.edu}}
\affiliation{Department of Physics and Astronomy, State University of New York at Stony Brook, \\ Stony Brook, NY 11794-3800, USA}

\author{Tuan K. Do}
\email{{ktdo@ucsb.edu}}
\affiliation{Department of Mathematics,  University of California, Santa Barbara, CA
93106, USA}

\author{Trung V. Phan}
\email{{tphan@natsci.claremont.edu}}
\affiliation{Department of Natural Sciences, Scripps and Pitzer Colleges, \\ Claremont Colleges Consortium, Claremont, CA 91711, USA}

\begin{abstract}
We introduce a quantum linear system solving algorithm based on the Kaczmarz method, a widely used workhorse for large linear systems and least-squares problems that updates the solution by enforcing one equation at a time. Its simplicity and low memory cost make it a practical choice across data regression, tomographic reconstruction, and optimization. In contrast to many existing quantum linear solvers, our method does not rely on oracle access to query entries, relaxing a key practicality bottleneck. In particular, when the rank of the system of interest is sufficiently small and the rows of the matrix of interest admit an appropriate structure, we achieve circuit complexity $\mathcal{O}\left(\frac{1}{\varepsilon}\log m\right)$, where $m$ is the number of variables and $\varepsilon$ is the target precision, without dependence on the sparsity $s$, and could possibly be without explicit dependence on condition number $\kappa$. This shows a significant improvement over previous quantum linear solvers where the dependence on $\kappa,s$ is at least linear. At the same time, when the rows have an arbitrary structure and have at most $s$ nonzero entries, we obtain the circuit depth $\mathcal{O}\left(\frac{1}{\varepsilon}\log s\right)$ using extra $\mathcal{O}(s)$ ancilla qubits, so the depth grows only logarithmically with sparsity $s$. When the sparsity $s$ grows as $\mathcal{O}(\log m)$, then our method can achieve an exponential improvement with respect to circuit depth compared to existing quantum algorithms, while using (asymptotically) the same amount of qubits. 
\end{abstract}

\maketitle

\section{Introduction}

Linear systems reside at the core of scientific computing, such as data regression/fitting \cite{strang2022introduction}, inverse problems such as tomographic reconstruction \cite{gordon1970algebraic,herman2009fundamentals}, and large-scale modern optimization \cite{wright1999numerical}. These tasks repeatedly boil down to solving systems of many linear algebraic equations --- often overdetermined and corrupted by noise --- that must be handled efficiently at scale. A powerful classical approach in this regime is the \textit{Kaczmarz method} \cite{kaczmarz1993approximate}, a lightweight ``row-action'' iteration that refines the estimate by enforcing one equation at a time, offering strong practical performance with minimal memory overhead in big data settings. As data sizes and model complexities continue to grow \cite{rydning2018digitization,kaplan2020scaling}, these linear-algebra subroutines increasingly dominate end-to-end pipelines, making it timely to explore quantum methods that can accelerate these fundamental linear-solve workloads across many applications.

Quantum algorithm for solving linear algebraic equations has stood out as one of the most promising quantum computer's application. The first quantum algorithm for solving sparse linear system was introduced in \cite{harrow2009quantum}, showing an exponential improvement compared to the best-known classical algorithm in relative to the dimension. In particular, it was shown in \cite{harrow2009quantum} that inverting a matrix is $\rm BQP$-complete, highlighting that it is unlikely for classical computers to match the quantum complexity. Subsequently, quantum linear solving algorithms have been refined and improved in a series of works. Ref.~\cite{childs2017quantum} introduced a quantum linear solvers with exponentially improved dependence on error tolerance. Ref.~\cite{clader2013preconditioned} introduced a namely preconditioned quantum linear solvers, which can handle ill-conditioned linear systems more efficiently than \cite{harrow2009quantum, childs2017quantum}. Ref.~\cite{huang2021near} outlined a variational quantum linear solvers, which was shown heuristically to perform well in practice. Ref.~\cite{wossnig2018quantum} constructed a quantum linear solvers suitable for dense system. Ref.~\cite{shao2020row} shares a certain similarity to ours, as both methods rely on row iterations. As we will show more explicitly later, our method gains certain advantage compared to \cite{shao2020row}. 

In this work, we develop a quantum version of the classical Kaczmarz algorithm. While quantum adaptations of Kaczmarz and related quantum linear solvers have been studied \cite{harrow2009quantum,childs2017quantum,wossnig2018quantum,clader2013preconditioned,huang2021near,shao2020row}, we advance beyond previous approaches on two fronts: we obtain \textit{faster scaling} in relevant regimes and, crucially, we adopt a substantially \textit{lighter input-access} model by avoiding QRAM/oracle queries of individual matrix entries --- thus the performance of our algorithm does not hinge on a data-access assumption that is likely to be the hardest part to realize in practice. Our paper is organized as follows. In Section \ref{sec: classicalalgorithm}, we review the classical Kaczmarz method, including its randomized iteration and convergence behavior. Section \ref{sec: quantumalgorithm} presents the quantum Kaczmarz algorithm: we state the input-access assumptions and state-preparation routines (Section \ref{sec: inputassumption}), construct the quantum row-action update via block-encodings then prove the main complexity bounds (Section \ref{sec: quantumkaczmarz}), and followed by a discussion and comparison with prior quantum linear solvers (Section \ref{sec: discussion}). The Appendices collect the required background on block-encoding/QSVT tools (Appendix \ref{sec: summaryofnecessarytechniques}) and provide the detailed complexity analysis underlying the main theorem (Appendix \ref{sec: analysis}).

\section{Classical Kaczmarz Method}
\label{sec: classicalalgorithm}
Consider a linear system
\begin{equation}
    Ax = b \ ,
\label{eq:lin_sys}
\end{equation}
where $A$ is a $n \times m$ real-valued matrix, $b$ is a known $n$ real-valued vector, and $x$ is the unknown $m$ real-valued vector the unknown vector to be estimated. For notational convenience, we write
$$b = \begin{pmatrix} b_1 \\ b_2 \\ ... \\ b_n \end{pmatrix} \in \mathbb{R}^n \ , \ x = \begin{pmatrix} x_1 \\ x_2 \\ ... \\ x_m \end{pmatrix} \in \mathbb{R}^m \ , \ $$
and represent the matrix $A$ row-wise as
$$A = \begin{pmatrix} a_1^\top \\ a_2^\top \\ ... \\ a_n^\top \end{pmatrix} \in \mathbb{R}^{n \times m} \ , \ a_j = \begin{pmatrix} a_{j1} \\ a_{j2} \\ ... \\ a_{jm} \end{pmatrix}  \in \mathbb{R}^m \ , $$
where $^\top$ is the transpose operation, so that the system Eq. \eqref{eq:lin_sys} is equivalent to the collection of $n$ scalar linear equations
$$a_j^\top x = a_{j1} x_1 + a_{j2} x_2 + ... + a_{jm} x_m = b_j \ . $$
In many applications, the matrix $A$ is \textit{not} required to be a square. Of particular interest are
\emph{rectangular} systems, in which
the number of equations is at least the number of unknowns, i.e. $n\ge m$. In this case, Eq. \eqref{eq:lin_sys} may have no
exact solution due to measurement noise or modeling error, and one often seeks
an \textit{approximate solution}.

A classical approach for solving this problem at large scale is the algebraic reconstruction technique known as the \textit{Kaczmarz method} \cite{kaczmarz1993approximate}, an
iterative row-action scheme that updates the current estimate by enforcing one
equation at a time \cite{censor1981row}. The procedure is as follows:
\begin{method}[Classical Kaczmarz method for solving linear system]
\label{algo: classicalkaczmarz}
Let $Ax = b$ be the linear system of interest where $A \in \Rbb^{n \times m}, b \in \Rbb^n$. 
\begin{itemize}
    \item \underline{Step 0:} We start with an initial guess $x^{(0)}$, often chosen as the zero-vector $(0,0,...,0)^\top$.
    \item \underline{Step $k\geq 1$:} Pick an index $j_k \in\{1,2,\ldots,n\}$, often cyclically or at random \cite{strohmer2009randomized}, and update the guess iteratively via:
    \begin{equation}
        x^{(k+1)} = x^{(k)} + \lambda \left( \frac{b_{j_k}-a_{j_k}^\top x^{(k)}}{||a_{j_k}||_2} \right) a_{j_k} \ ,
        \label{eq:kaczmarz_update}
    \end{equation}
    where $||a_{j_k}||_2 =a_{j_k}^\top a_{j_k} $ is the row $l_2$-norm and $\lambda$ is the relaxation parameter, which is often chosen between $0$ and $2$. For $\lambda=1$, this update can be interpreted as an \textit{exact projection}\footnote{When $\lambda=1$, the update is the orthogonal projection of $x^{(k-1)}$ onto the hyperplane $a_{j_k}^\top x = b_{j_k}$, as it produces an iterate $x^{(k)}$ satisfies $a_{j_k}^\top x^{(k)} = b_{j_k}$ exactly.}.
\end{itemize}
\end{method}

In other words, the Kaczmarz method views each row-equation $a_j^\top x=b_j$ as a hyperplane in
$\mathbb{R}^m$, then generates a sequence $\{x^{(k)}\}$ by repeatedly
projecting onto these hyperplanes by applying Eq. \eqref{eq:kaczmarz_update}. The iteration is terminated after a prescribed number of steps, or once a chosen convergence criterion is met. It has been shown rigorously that, with randomized row selection (sampled with probability proportional to the row $l_2$-norms), the relative residual with respect to the least-squares solution is guaranteed to converge at an exponential rate \cite{strohmer2009randomized}. Specifically, it was shown that if the row is chosen randomly with probability:
\begin{align}
    \mathbb{P}(j_k = j) = \frac{||a_j||_2^2}{||A||_F} \ ,
\end{align}
then the expected deviation at iteration $k \gg 1$ decreases gemetrically as:
\begin{equation}
\begin{split}
    &\mathbb{E}\left(  || x^{(k)} - x_{\rm sol} ||^2 \right)
    \\
    & \ \ \ \ \leq \left( 1 - \frac{\sigma^2_{\min}(A)}{||A||_F^2} \right)^k  || x^{(0)} - x_{\rm sol}||^2_2 \ .
\end{split}
\end{equation}
where $\sigma_{\min}(A)$ is the smallest singular value of $A$ (in magnitude), $||A||_F$ is the Frobenius norm of $A$, and $x_{\rm sol}$ is the true solution (if there is any) to the original linear system $Ax = b$. It can be clearly deduced from the above inequality that for a desired additive error $\epsilon$, by setting the right-hand equal to $\epsilon$, the number of iteration steps $T$ (i.e., $T=\max k$) needs to be: 
\begin{equation}
    T = \mathcal{O}\left( \frac{1}{ -\log \big(1 - \frac{\sigma^2_{\min}(A)}{||A||_F^2}  \big)}\log \frac{1}{\epsilon } \right) \ .
\label{iter_time}
\end{equation}
Compared to more traditional approaches, such as direct inversion \cite{golub2013matrix}, a drawback of the Kaczmarz method is that it is iterative and typically approaches the solution only \textit{asymptotically} rather than reaching it exactly in finitely many steps. Its main advantages are simplicity and low per-iteration cost, which often produces a reasonable approximate solution in the \textit{least-squares} sense very quickly.

\section{Quantum Kaczmarz Method}
\label{sec: quantumalgorithm}

In this section, we outline the construction of the quantum Kaczmarz method for solving linear systems of the form $Ax = b$ where $A,b$ is the matrix and the vector of the forms discussed above. Our quantum algorithm is built on the classical version described in Algo.~\ref{algo: classicalkaczmarz}. In the following, we first describe some input assumptions that our quantum algorithm requires. Then, we construct a quantum procedure for carrying out the iteration of the (classical) Kaczmarz algorithm. 

\subsection{Input assumption}
\label{sec: inputassumption}
The information that we assume to have in this work are:
\begin{itemize}
    \item The matrix $A$ has operator norm $||A|| \leq 1$ (i.e., its largest singular value is less than $1$). We also assume the vector $b$ has $l_2$-norm $||b||_{2}\leq 1$. 
    \item Classical knowledge of the entries $\{ b_j\}_{j=1}^n$ of vector $b$.
    \item Classical knowledge of the entries $\{  a_{jl}\}_{j,l=1}^{m,n}$ of matrix $A$. 
    \item $l_2$-norms $\{  ||a_j||_2 \}_{j=1}^n$ of the rows of $A$. 
\end{itemize}
We remark that the first assumption regarding the operator norm of $A$ and $l_2$-norm of $b$ is without loss of generalization, as we can always rescale the system by some constant factor. In fact, this assumption also appears in all related works \cite{harrow2009quantum, childs2017quantum, wossnig2018quantum}. The next three assumptions, particularly the classical knowledge of the entries of matrix $A$ plus the $l_2$-norm of rows of $A$, allow us to leverage many recent advances in quantum state preparation protocols, e.g., \cite{zhang2022quantum, mcardle2022quantum, marin2023quantum, nakaji2022approximate, zoufal2019quantum}, to (approximately) prepare the state $\{ \ket{a_j}\}_{j=1}^n$ (where for each $j$, $\ket{a_j}\equiv \frac{a_j}{||a_j||_2}$). A concrete procedure for state preparation can be found in these references, and we refer the interested readers to them. Here, for our purpose, we recapitulate their results in the following lemma:
\begin{lemma}
\label{lemma: statepreparation}
    Provided the classical knowledge of $\{  a_{ij}\}_{i,j=1}^{m,n}$ and $l_2$-norms $\{  ||a_j||_2 \}_{j=1}^n$ of the rows $a_j$ of $A$, for each $j$, if $a_j$ admits the structure as in \cite{mcardle2022quantum, marin2023quantum, nakaji2022approximate, zoufal2019quantum}, the state $\ket{a_j}$ can be prepared using a quantum circuit of gate complexity $\mathcal{O}\left( \log m\right)$, plus $\mathcal{O}(1)$ ancilla qubits. In particular, for a generally arbitrary structure of $a_j$, the state $\ket{a_j}$ can be prepared using a quantum circuit of depth $\mathcal{O}\left( \log s_j \right)$ at the trade-off of using $\mathcal{O}(s_j)$ ancilla qubits (see Ref.~\cite{zhang2022quantum}), where $s_j$ is the number of nonzero entries of the row $a_j$ of $A$. 
\end{lemma}
From this recipe, we now proceed to describe our quantum Kaczmarz method for solving linear systems. 

\subsection{Quantum Kaczmarz algorithm for solving linear equations}
\label{sec: quantumkaczmarz}

For every iteration $k$, the algorithm proceeds through a fixed sequence of five steps, which we describe in detail below:

\underline{Step $1$:} The first goal is to prepare the block-encoding of the factor $a^\top_{j_k} x^{(k)}$ (by this we mean a unitary that has the top-left entry, or the entry in the first row and column, to be $a^\top_{j_k} x^{(k)} $). This can be done as follows. According to Lemma \ref{lemma: statepreparation}, at the $k$-th iteration step $j_k$ (for $j_k \in [1,2,...,n]$), we can construct a unitary, denoted by $U_{j_k}$, which could prepare the state $\ket{a_{j_k}}$. It can also be seen that the first column of the unitary $U_{j_k} $ is $\ket{a_{j_k}} $. Suppose further that at this $k$-th iteration step, we have obtained a unitary $U_{x^{(k)}}$ which is a block-encoding of a matrix containing the temporal solution $x^{(k)}$ in the first column. Then we can use Lemma \ref{lemma: product} to construct the block-encoding of $U_{j_k}^\dagger U_{x^{(k)}}$. Because the first column of $ U_{j_k}$ is $ \ket{a_{j_k}} $, the first row of $ U_{j_k}^\dagger $ is $\ket{a_{j_k}}  $. So, the top-left entry (the one in the first row and first column) of $ U_{j_k}^\dagger U_{x^{(k)}}$ is  $\bra{a_{j_k}}x^{(k)}  $. 

After this, we can then use Lemma \ref{lemma: scale} with the scaling factor $||a_{j_k}||_2$ to obtain the block-encoding of 
$$ ||a_{j_k}||_2  U_{j_k}^\dagger U_{x^{(k)}} \ , $$
which contains the product $ ||a_{j_k}||_2  \bra{a_{j_k}}x^{(k)} =  a^\top_{j_k} x^{(k)} $ in the first row and column entry. We note that we have used the following property: $$||a_{j_k}||_2  \bra{a_{j_k}} = ||a_{j_k}||_2  \frac{a_{j_k}}{||a_{j_k}||_2}  = a_{j_k} \ . $$ 

\underline{Step $2$}: Now we need the block-encoding of $ b_{j_k} $, which can be simply obtained by considering a rotation gate: 
\begin{align}
    R_x(\theta) = \begin{pmatrix}
    \cos( \theta/2) & -i \sin(\theta/2) \\
    -i \sin(\theta/2) & \cos(\theta/2)
    \end{pmatrix} \ .
\end{align}
By choosing $\theta$ so that $\cos(\theta/2) = b_{j_k}$ (which is always possible because we have assumed that the $l_2$-norm $||b||_2 \leq 1$ so all entries $\{b_j\}_{j=1}^n$ are less than 1), we obtain the unitary $R_x(\theta)$ that contains $b_{j_k}$ as the top-left entry. 

\underline{Step $3$}: Next, given that the identity matrix $\Ibb$ of arbitrary dimension can be trivially obtained (and also block-encoded, see Def \ref{def: blockencode}), we use Lemma \ref{lemma: tensorproduct} to construct the block-encoding of $ R_x(\theta) \otimes \Ibb$ where the dimension of $\Ibb$ is chosen so that the dimension of $ R_x(\theta) \otimes \Ibb$ matches the dimension of $||a_{j_k}||_2U_{j_k}^\dagger U_{x^{(k)}} $. 

From these block-encodings, we can use Lemma \ref{lemma: sumencoding} to construct the block-encoding of their subtraction, e.g., 
$$ \frac12 \hat{\Xi} \ \  \text{with} \ \  \hat{\Xi} \equiv  R_x(\theta) \otimes \Ibb - ||a_{j_k}||_2 U_{j_k}^\dagger U_{x^{(k)}} \ .$$
This operator has the top-left entry to be $ \frac{1}{2} \left( b_{j_k} - a^\top_{j_k} x^{(k)}   \right)$. 

From the block-encoding above, we then use Lemma  \ref{lemma: tensorproduct} again to construct the block-encoding of 
$$ \frac12 \hat{\Xi} \otimes U_{j_k} \ . $$
The first column of this operator can be seen as $\frac{1}{2} \left( b_{j_k} - a^\top_{j_k} x^{(k)}   \right) \ket{a_{j_k}} $, as the first column of $ U_{j_k}$ is $\ket{a_{j_j}}$. Then we can use Lemma \ref{lemma: amp_amp} to multiply the above block-encoded operator with a factor $\frac{1}{||a_{j_k}||_2}$, so we obtain the block-encoding of:
\begin{align}
    \frac{1}{2||a_{j_k}||_2}\hat{\Xi} \otimes U_{j_k}
\end{align}
From this block-encoding, we can use Lemma \ref{lemma: scale} with some chosen scaling factor $\lambda$, to obtain the block-encoding of:
\begin{align}
    \frac{\lambda}{2||a_{j_k}||_2}\hat{\Xi} \otimes U_{j_k}
\end{align}

\underline{Step $4$}: From this block-encoding and also the block-encoding $U_{x^{(k)}}$, Lemma \ref{lemma: sumencoding} allows us to obtain the block-encoding of:
\begin{align}
    \frac{1}{2} \left(  U_{x^{(k)}} -\frac{\lambda}{2||a_{j_k}||_2}\hat{\Xi} \otimes U_{j_k}\right) \ .
    \label{iii2}
\end{align}
Recall that $U_{x^{(k)}}$ is the block-encoding of a matrix that has $x^{(k)}$ as the first column. So the above operator is, in fact, block-encoding of a matrix that has the following vector as the first column:
\begin{align}
    \frac{1}{2}\left[  x^{(k)} - \frac{\lambda}{2||a_{j_k}||_2} \left( b_{j_k} - a^\top_{j_k} x^{(k)}   \right) \ket{a_{j_k}}  \right] \ .
\end{align}
By interpreting the prefactor $\lambda/2$ as corresponding to a reasonable selection of the relaxation parameter $\lambda$ (by a slight abuse of notation), and note that $ \ket{a_{j_k}}  = \frac{a_{j_k}}{||a_{j_k}||_2}$, the above vector can be identified as:
\begin{align}
    \frac{1}{2}\left[  x^{(k)} -  \lambda  \left( b_{j_k} - a^\top_{j_k} x^{(k)} \frac{a_{j_k}}{||a_{j_k}||_2^2} \right)   \right] \ .
\end{align}
According to the second step of Algo.~\ref{algo: classicalkaczmarz}, this vector is equivalent to $\frac{1}{2} x^{(k+1)}$. To sum up, starting from the unitary $U_{x^{(k)}}$ which block-encodes a matrix having $x^{(k)}$ as the first column, and the unitary $U_{j_k}$ which has $\ket{a_{j_k}}$ as the first column, we have shown how to obtain the block-encoding of an operator (Eqn.~\ref{iii2}) having $\frac{1}{2} x^{(k+1)} $ as the first column. 

\underline{Step $5$}: The factor $1/2$ can be removed by using Lemma \ref{lemma: amp_amp} to multiply the block-encoded operator in Eqn.~\ref{iii2} with a factor of $2$, resulting in the unitary $U_{x^{(k+1)}}$, which block-encodes the operator:
\begin{align}
     U_{x^{(k)}} - \frac{1}{2} \hat{\Xi} \otimes U_{j_k}  \ .
\end{align}
As indicated above, this operator has $ x^{(k)}$ as the first column. With this unitary block-encoding $U_{x^{(k+1)}}$, we can execute a similar procedure as above, albeit with $U_{x^{(k)}}$ replaced by $U_{ x^{(k+1)}}$. The outcome of this procedure is the unitary $U_{x^{(k+2)}}$ which block-encodes a matrix having $x^{(k+2)}$ as the first column. The procedure is then repeated using this new unitary $U_{x^{(k+2)}}$ in replacement of $U_{x^{(k+1)}}$ from the previous step. 

Recall from Algo.~\ref{algo: classicalkaczmarz} that the (classical) Kaczmarz method begins with an initial guess $x^{(0)}$. Without loss of generality, we choose an arbitrary state $\ket{x^{(0)}}$ with the known unitary preparation $U_{x^{(0)}}$, e.g., 
$$U_{x^{(0)}} \ket{0}^{\otimes \log m} = \ket{x^{(0)}} \ . $$ 
It can be seen that the first column of $U_{x^{(0)}}$ is $\ket{x^{(0)}}$. For a chosen $T$ and $U_{x^{(0)}}$ as the starting unitary, we can iterate the procedure outlined in the previous paragraphs $T$ times, obtaining 
$$U_{x^{(0)}}, U_{x^{(1)}}, U_{x^{(2)}}, ...., U_{x^{(T)}}$$
which block-encode the matrices having the temporal solutions $\{x^{k} \}_{k=1}^T$ as the first columns. In order to obtain the state $\ket{x^{(T)}} = \frac{x^{(T)}}{||x^{(T)}||_2}$  that corresponds to the (approximate) solution of the linear system, we can perform the following step. Taking the unitary $U_{x^{(T)}} $ and apply it to the state $\ket{\bf 0} \ket{0}^{\otimes \log m}$ (where $\ket{\bf 0}$ correspond to those ancilla qubits required to block-encode $U_{x^{(T)}} $). According to Def.~\ref{def: blockencode}, we obtain the following state:
\begin{align}
U_{x^{(T)}} \ket{\bf 0} \ket{0}^{\otimes \log m} = \ket{\bf 0} x^{(T)} + \ket{\rm Garbage} \ ,
\label{measurement}
\end{align}
where $ \ket{\rm Garbage}$ is some redundant state that is orthogonal to $ \ket{\bf 0} x^{(T)}$. By measuring the ancilla qubits and post-select on seeing $\ket{\bf 0}$, we can obtain the state $\ket{x^{(T)}}= \frac{x^{(T)}}{||x^{(T)}||_2}$. Thus, we have completed the quantum Kaczmarz algorithm for solving linear equations. 

For completeness, we summarize the whole procedure outlined above in the following:
\begin{method}[Quantum Kaczmarz method for solving linear system]
\label{algo: quantumkaczmarz}
    Let $Ax = b $ the linear system of interest, with $A,b$ admit four assumptions mentioned earlier (see Sec.~\ref{sec: inputassumption}). Let $U_{x^{(0)}}$ be the state preparation unitary with the first column to be $x^{ (0)}$ and fix $T$ to be the total iteration steps. Suppose that at $k$-th iteration step, we have obtained the unitary $U_{x^{(k)}}$ which block-encodes a matrix having the temporal solution $x^{ (k) }$ in the first column. Then iterate the following five-step procedure $T$ times:
    \begin{enumerate}
        \item Obtain the unitary block-encoding of an operator having top-left entry to be $a_{j_k}^T x^{(k)}$. 
        \item Obtain the unitary block-encoding of an operator having top-left entry to be $b_{j_k}$.
        \item Obtain the unitary block-encoding of an operator having the first column to be $\frac{1}{2} \left( b_{j_k} - a^\top_{j_k} x^{(k)}   \right) \ket{a_{j_k}}  $.
        \item Obtain the unitary block-encoding of an operator having the first column to be $ \frac{1}{2}x^{(k+1)}$. 
        \item Obtain the unitary block-encoding of an operator having the first column to be $ x^{(k+1)}$.
    \end{enumerate}
    As a final step after $T$ iterations of the above procedure, we apply the resultant unitary $U_{x^{(T)}}$ to the state $\ket{\bf 0}\ket{0}^{\otimes \log m}$, performing measurement on the ancilla system and post-select on seeing $\ket{\bf 0}$. 

    \noindent
    \textbf{ Output:} Quantum state $\ket{x^{(T)}}$ corresponds to $x^{(T)}$ which is an approximation to the solution of linear system $Ax = b$. 
\end{method} 
In Section \ref{sec: classicalalgorithm}, we have pointed out that for an additive precision $\epsilon$, i.e., $|| x_T - x_{\rm solution} || \leq \epsilon$, the necessary number of iterations $T$ is given by Eq. \eqref{iter_time}. An in-depth analysis of the quantum algorithm outlined above will be provided in the Appendix \ref{sec: analysis}. Here, for brevity, we recapitulate the main result in the following theorem:
\begin{theorem}
\label{theorem: mainresult}
Let the linear system of interest be $Ax = b$ for $A \in \Rbb^{n \times m}, b \in \Rbb^n$ with the four assumptions as stated earlier and $r_A$ denotes the rank of $A$. For an additive precision $\epsilon$, the Algorithm \ref{algo: quantumkaczmarz} can output $\ket{x_T} \equiv \frac{x_T}{||x_T||_2}$ such that $ || x_T - x_{\rm solution} || \leq \epsilon$. 
\begin{itemize}
    \item If for each $j$, the row $a_j$ admits the structure as indicated in \cite{nakaji2022approximate, zoufal2019quantum, marin2023quantum, mcardle2022quantum}, then the circuit complexity of the Algorithm \ref{algo: quantumkaczmarz} is $ \mathcal{O}\left( 2^{r_A} ||x_T||_2 \frac{1}{\epsilon} \log m   \right)$, with a total of extra $\mathcal{O}(1)$ ancilla qubits. 
\item For a general structure of $a_j$, let $s_j$ denotes the number of nonzero entries of $a_j$. Defining $s = \max \{  s_j \}_{j=1}^n$. Then the quantum circuit depth employed Algorithm \ref{algo: quantumkaczmarz} is $\mathcal{O}\left( 2^{r_A}||x_T||_2   \frac{1}{\epsilon} \log s\right)$, at the cost of requiring an extra $\mathcal{O}(s)$ ancilla qubits. 
\end{itemize}
\end{theorem}

\subsection{Discussion}
\label{sec: discussion}
In this section, we discuss our quantum algorithm from a broader perspective by examining its regime of best efficiency, as well as its potential advantage in relative to existing algorithms. \\

\noindent
\textbf{Dealing with rectangular linear system.} Our quantum algorithm is built on the classical Kaczmarz method, which is naturally well-suit for handling rectangular linear system. In this case, the unique solution might not exist, and thus, the quantum linear solvers in \cite{harrow2009quantum, childs2017quantum, wossnig2018quantum, clader2013preconditioned} are not able to find the solution, as in this case, the inverse of $A$ is ill-defined. At the same time, our quantum algorithm can still return the ``solution'', which in this case is understood from the perspective of lease-square, e.g., find $x$ that minimizes $||Ax - b||_2$. We point out that other quantum linear solving algorithms existing, which could also deal with rectangular case \cite{huang2021near, shao2020row}.\\

Another important factor in the complexity (Theorem \ref{theorem: mainresult}) is the exponential scaling on $r_A$ -- which is the rank of $A$. Due to this, our algorithm is most efficient when $r_A$ is sufficiently small, or that the rank of $A$ is sufficiently small. We point out that, as also noted in \cite{harrow2009quantum, childs2017quantum}, this is the regime in which their algorithms are not effective. Therefore, it suggests that our quantum algorithm can complement very well to the existing quantum linear solvers in the regime of rectangular system, where the number of equations $n$ exceed that of the number of unknowns.

\noindent
\textbf{Improvement over sparsity parameter $s$ and condition number $\kappa$.} In the case of square linear system, we set $n=m$ to be the primary dimension. To compare, we provide the table summarizing the complexity of existing quantum linear solvers. 
\begin{table}[htbp]
    \centering
    \begin{tabular}{|c|c|c|}
    \hline
      & Complexity & QRAM/Oracle \\
      \hline
        Ref.~\cite{harrow2009quantum} & $\mathcal{O} \left( \frac{1}{\epsilon}s^2 \kappa \log m  \right)$ &  YES\\
        \hline
        Ref.~\cite{childs2017quantum} & $\mathcal{O}\left(  s\kappa^2  \log \frac{m}{\epsilon   } \right)$  &  YES \\
        \hline
        Ref.~\cite{wossnig2018quantum} &  $\mathcal{O}\left( \kappa^2 \sqrt{m} \ \rm polylog \frac{m}{\epsilon}  \right)$ & YES \\
        \hline
        Ref.~\cite{clader2013preconditioned}&  $\mathcal{O}\left( s^7 \frac{1}{\epsilon} \log m \right)$   & YES \\
        \hline
        Ref.~\cite{huang2021near} &  Heuristic  &  NO\\
        \hline
        Ref.~\cite{shao2020row} & $ \mathcal{O}\left(  \kappa^2  \log \frac{1}{\epsilon}\log m\right)$ &  YES \\
        \hline 
        Our work & $\mathcal{O}\left(  \frac{1}{\epsilon}\log m\right)$ or $\mathcal{O}\left( \frac{1}{\epsilon}\log s \right)$ & NO \\
\hline
    \end{tabular}
    \caption{Table summarizing the circuit complexities of existing quantum linear solving algorithms. In the above table, $\kappa$ denotes the condition number of $A$, and $s$ denotes the sparsity of $A$ (the maximum number of nonzero entries in each row or column). } 
    \label{tab: comparison}
\end{table}

From Table \ref{tab: comparison}, we can see that our method exhibits a certain advantage with respect to the condition number $\kappa$ and the sparsity parameter $s$. More specifically, as indicated in Thm.~\ref{theorem: mainresult}, in the case where all rows of $A$ admit certain structures, then our complexity would be $\mathcal{O}\left( ||x_T||_2  \frac{1}{\epsilon}\log m \right)$, which is independent of $s$. We remark that this complexity can also be independent of $\kappa$ if $||x_T||_2 = \mathcal{O}(1)$. In the Appendix \ref{sec: analysis}, it will be shown (using result of \cite{harrow2009quantum}) that $||x_T||_2$ is upper bounded by $\mathcal{O}(\kappa)$. So in the worst case, our complexity has linear scaling on $\kappa$. Sill, this is a major improvement over existing results where the dependence on $\kappa,s$ is linear to polynomial. At the same time, if the rows of $A$ have an arbitrary structure, then our quantum algorithm can achieve a circuit of depth logarithmical in the sparsity parameter $s$, at the cost of using $\mathcal{O}(s)$ extra ancilla qubits. Therefore, in practice, this is only qubit-efficient when $s$ scales polylogarithmically in the dimension $m$. In this case, our algorithm achieves exponential improvement (in terms of circuit depth) with respect to the sparsity $s$ over existing works. The number of ancilla qubits required is $\mathcal{O}(s) = \mathcal{O}(\log n)$, so the total number of qubits is $\mathcal{O}(\log n)$, which is similar to existing works.  \\

\noindent
\textbf{Relaxation over strong input assumption.} As also indicated in Table \ref{tab: comparison}, most of existing quantum linear solving algorithms assume the access to an oracle which could efficiently query the entries of $A$. A few proposals have been made to realize this oracle, for example, quantum random access memory (QRAM) \cite{giovannetti2008architectures, giovannetti2008quantum}. However, this oracle assumption has been deemed a fairly strong input assumptions. On one hand, large scale and fault-tolerance QRAM is yet available, making the quantum algorithms which rely on QRAM difficult to experimentally realize. On the other hand, progress on dequantization algorithm \cite{tang2019quantum, gilyen2018quantum, shao2022faster} have revealed that without the oracle assumption, quantum algorithms cannot achieve exponential speedup, at least in general setting. In the same works \cite{tang2019quantum, gilyen2018quantum, shao2022faster}, the authors specifically show that if classical algorithms have access to a particular input, which is analogous to the oracle assumption, then classical computers can solve many tasks with polylogarithmical complexity. \\

\noindent
\textbf{Trade-off over inverse of error tolerance.} From table \ref{tab: comparison}, it can be seen that our method has linear scaling in $\frac{1}{\epsilon}$, which is exponentially less efficient than most existing works (except \cite{clader2013preconditioned}). This stems from the fact that our algorithm's complexity admits exponential scaling on $T$, as our method is iterative and in each step we need to employ the outcome from the previous step multiple times. We regard this is a trade-off for a better dependence on $s$ and possibly $\kappa$ (if $||x_T||_2$ behaves as $\mathcal{O}(1)$), and also that our method does not depend on oracle assumption. Given this, we believe that in reality, our method can be a nice complementary to existing quantum linear solvers. For those linear system with large condition number and sparsity, or when the oracle access is not efficient to realize, our algorithm can be more capable.

\section{Conclusion}
\label{sec: conclusion}
In this work, we have outlined a quantum Kaczmarz algorithm for solving linear algebraic system. Our algorithm is directly built on the classical Kaczmarz algorithm, which solves the linear system by iteratively updating the solution based on random column selection. Upon appropriate input assumption regarding the structure of $A,b$, as well as block-encoded operator containing the temporal solution $x^{(k)}$, we have shown how to construct the block-encoding of $a_{j_k}^T x^{(k)}, b_{j_k}, \frac{1}{2}( b_{j_k} - a_{j_k}^T x^{(k)}) $ and finally of an operator having the desired updated solution $x^{(k+1)}$. The procedure is then iterated for a total of $T$ times, followed by an application to a known state and measurement. The outcome of the post-selected measurement is the desired approximation to the solution of the given linear system. We then provide a discussion, showing that our algorithm can be advantageous compared to prior quantum linear solvers in certain aspects. Despite having major improvement on $\kappa,s$, our method turns out to have exponential scaling on $T$, which leads to a linear scaling on $\frac{1}{\epsilon}$. This is exponentially less efficient than \cite{harrow2009quantum, childs2017quantum}. We regard this is a reasonable trade-off for the improvement on $\kappa,s$. Yet, it is not known to us if this trade-off is a must. Therefore, how to improve this exponential scaling on $T$ (and hence on $\frac{1}{\epsilon}$) is an interesting avenue.

\section*{Acknowledgments}
This work was supported by the U.S. Department of Energy, Office of Science, National Quantum Information Science Research Centers, Co-design Center for Quantum Advantage (C2QA) under Contract No. DE-SC0012704. N.A.N. also acknowledge support from the Center for Distributed Quantum Processing at Stony Brook University. N.A.N. thanks the hospitality of Harvard University where he has an academic visit during the completion of this project.

\bibliography{ref.bib}
\bibliographystyle{unsrt}

\appendix
\onecolumngrid

\section{Block-encoding and quantum singular value transformation}
\label{sec: summaryofnecessarytechniques}
We briefly summarize the essential quantum tools used in our algorithm. For conciseness, we highlight only the main results and omit technical details, which are thoroughly covered in~\cite{gilyen2019quantum}. An identical summary is also presented in~\cite{lee2025new}.

\begin{definition}[Block-encoding unitary, see e.g.~\cite{low2017optimal, low2019hamiltonian, gilyen2019quantum}]
\label{def: blockencode} 
Let $A$ be a Hermitian matrix of size $N \times N$ with operator norm $\norm{A} < 1$. A unitary matrix $U$ is said to be an \emph{exact block encoding} of $A$ if
\begin{align}
    U = \begin{pmatrix}
       A & * \\
       * & * \\
    \end{pmatrix},
\end{align}
where the top-left block of $U$ corresponds to $A$. Equivalently, one can write
\begin{equation}
    U = \ket{\mathbf{0}}\bra{\mathbf{0}} \otimes A + (\cdots),    
\end{equation}
where $\ket{\mathbf{0}}$ denotes an ancillary state used for block encoding, and $(\cdots)$ represents the remaining components orthogonal to $\ket{\mathbf{0}}\bra{\mathbf{0}} \otimes A$. If instead $U$ satisfies
\begin{equation}
    U = \ket{\mathbf{0}}\bra{\mathbf{0}} \otimes \tilde{A} + (\cdots),
\end{equation}
for some $\tilde{A}$ such that $\|\tilde{A} - A\| \leq \epsilon$, then $U$ is called an {$\epsilon$-approximate block encoding} of $A$. Furthermore, the action of $U$ on a state $\ket{\mathbf{0}}\ket{\phi}$ is given by
\begin{align}
    \label{eqn: action}
    U \ket{\mathbf{0}}\ket{\phi} = \ket{\mathbf{0}} A\ket{\phi} + \ket{\mathrm{Garbage}},
\end{align}
where $\ket{\mathrm{Garbage}}$ is a state orthogonal to $\ket{\mathbf{0}}A\ket{\phi}$. The circuit complexity (e.g., depth) of $U$ is referred to as the {complexity of block encoding $A$}.
\end{definition}

\begin{lemma}[Amplification, Theorem 30 of~\cite{gilyen2019quantum}]
\label{lemma: amp_amp}
Let $U$, $\Pi$, $\widetilde{\Pi} \in {\rm End}(\mathcal{H}_U)$ be linear operators on $\mathcal{H}_U$ such that $U$ is a unitary, and $\Pi$, $\widetilde{\Pi}$ are orthogonal projectors. 
Let $\gamma>1$ and $\delta,\epsilon \in (0,\frac{1}{2})$. 
Suppose that $\widetilde{\Pi}U\Pi=W \Sigma V^\dagger=\sum_{i}\varsigma_i\ket{w_i}\bra{v_i}$ is a singular value decomposition. 
Then there is an $m= \mathcal{O} \Big(\frac{\gamma}{\delta}
\log \left(\frac{\gamma}{\epsilon} \right)\Big)$ and an efficiently computable $\Phi\in\mathbb{R}^m$ such that
\begin{align}
\left(\bra{+}\otimes\widetilde{\Pi}_{\leq\frac{1-\delta}{\gamma}}\right)U_\Phi \left(\ket{+}\otimes\Pi_{\leq\frac{1-\delta}{\gamma}}\right) = \sum_{i\colon\varsigma_i\leq \frac{1- \delta}{\gamma} }\tilde{\varsigma}_i\ket{w_i}\bra{v_i} , \text{ where } \Big|\!\Big|\frac{\tilde{\varsigma}_i}{\gamma\varsigma_i}-1 \Big|\!\Big|\leq \epsilon.
\end{align}
Moreover, $U_\Phi$ can be implemented using a single ancilla qubit with $m$ uses of $U$ and $U^\dagger$, $m$ uses of C$_\Pi$NOT and $m$ uses of C$_{\widetilde{\Pi}}$NOT gates and $m$ single qubit gates.
Here,
\begin{itemize}
\item C$_\Pi$NOT$:=X \otimes \Pi + I \otimes (I - \Pi)$ and a similar definition for C$_{\widetilde{\Pi}}$NOT; see Definition 2 in \cite{gilyen2019quantum},
\item $U_\Phi$: alternating phase modulation sequence; see Definition 15 in \cite{gilyen2019quantum},
\item $\Pi_{\leq \delta}$, $\widetilde{\Pi}_{\leq \delta}$: singular value threshold projectors; see Definition 24 in \cite{gilyen2019quantum}.
\end{itemize}
\end{lemma}
Based on~\ref{def: blockencode}, several properties, though immediate, are of particular importance and are listed below.
\begin{remark}[Properties of block-encoding unitary]
The block-encoding framework has the following immediate consequences:
\begin{enumerate}[label=(\roman*)]
    \item Any unitary $U$ is trivially an exact block encoding of itself.
    \item If $U$ is a block encoding of $A$, then so is $\Ibb_m \otimes U$ for any $m \geq 1$.
    \item The identity matrix $\Ibb_m$ can be trivially block encoded, for example, by $\sigma_z \otimes \Ibb_m$.
\end{enumerate}
\end{remark}

Given a set of block-encoded operators, various arithmetic operations can be done with them. Here, we simply introduce some key operations that are especially relevant to our algorithm, focusing on how they are implemented and their time complexity, without going into proofs. For more detailed explanations, see~\cite{gilyen2019quantum, camps2020approximate}.

\begin{lemma}[Informal, product of block-encoded operators, see e.g.~\cite{gilyen2019quantum}]
\label{lemma: product}
    Given unitary block encodings of two matrices $A_1$ and $A_2$, with respective implementation complexities $T_1$ and $T_2$, there exists an efficient procedure for constructing a unitary block encoding of the product $A_1 A_2$ with complexity $T_1 + T_2$.
\end{lemma}

\begin{lemma}[Informal, tensor product of block-encoded operators, see e.g.~{\cite[Theorem 1]{camps2020approximate}}]\label{lemma: tensorproduct}
    Given unitary block-encodings $\{U_i\}_{i=1}^m$ of multiple operators $\{M_i\}_{i=1}^m$ (assumed to be exact), there exists a procedure that constructs a unitary block-encoding of $\bigotimes_{i=1}^m M_i$ using a single application of each $U_i$ and $\mathcal{O}(1)$ SWAP gates.
\end{lemma}

\begin{lemma}[Informal, linear combination of block-encoded operators, see e.g.~{\cite[Theorem 52]{gilyen2019quantum}}]
    Given the unitary block encoding of multiple operators $\{A_i\}_{i=1}^m$. Then, there is a procedure that produces a unitary block encoding operator of $\sum_{i=1}^m \pm (A_i/m) $ in time complexity $\mathcal{O}(m)$, e.g., using the block encoding of each operator $A_i$ a single time. 
    \label{lemma: sumencoding}
\end{lemma}

\begin{lemma}[Informal, Scaling multiplication of block-encoded operators] 
\label{lemma: scale}
    Given a block encoding of some matrix $A$, as in~\ref{def: blockencode}, the block encoding of $A/p$ where $p > 1$ can be prepared with an extra $\mathcal{O}(1)$ cost.
\end{lemma}



\begin{lemma}[Matrix inversion, see e.g.,~\cite{gilyen2019quantum, childs2017quantum}]\label{lemma: matrixinversion}
Given a block encoding of some matrix $A$  with operator norm $||A|| \leq 1$ and block-encoding complexity $T_A$, then there is a quantum circuit producing an $\epsilon$-approximated block encoding of ${A^{-1}}/{\kappa}$ where $\kappa$ is the conditional number of $A$. The complexity of this quantum circuit is $\mathcal{O}\left( \kappa T_A \log \left({1}/{\epsilon}\right)\right)$. 
\end{lemma}

\section{Upper bound and lower bound on $T$}
In Section \ref{sec: classicalalgorithm}, we have pointed out that for an additive precision $\epsilon$, the number of iterations needs to be (we replace $k$ by $T$ to match the notation we have used this section):
$$T  = \mathcal{O}\left( \frac{1}{ -\log \big(1 - \frac{\sigma^2_{\min}(A)}{||A||_F^2}  \big)}\log \frac{1}{\epsilon } \right)$$
It is of known property that the Frobenius norm $||A||_F^2$ is the sum of squared of singular values of $A$. So it holds that:
\begin{align}
    ||A||_F^2 \leq r_A \sigma^2_{\max}(A)
\end{align}
where $r_A$ is the rank of $A$. Therefore, it holds that:
\begin{align}
 \frac{\sigma^2_{\min}(A)}{||A||_F^2} \geq \frac{ \sigma^2_{\min}(A)}{r_A \sigma^2_{\max}(A)}
\end{align}
which leads to:
\begin{align}
    1- \frac{\sigma^2_{\min}(A)}{||A||_F^2} \leq 1- \frac{1}{r_A \kappa_A^2} 
\end{align}
where $\kappa_A \equiv \frac{\sigma_{\max}(A)}{\sigma_{\min}(A)}$ is the condition number of $A$. Thus:
\begin{align}
    -\log \big(1 - \frac{\sigma^2_{\min}(A)}{||A||_F^2}  \big) \geq \log \frac{r_A \kappa_A^2}{r_A \kappa_A^2-1} \\
    \longrightarrow \frac{1}{ -\log \big(1 - \frac{\sigma^2_{\min}(A)}{||A||_F^2}  \big)} \leq \frac{1}{ \log \frac{r_A \kappa_A^2}{r_A \kappa_A^2-1} }
\end{align}
Therefore, the number of iterations $T$ is bounded by:
\begin{align}
    T = \mathcal{O}\left( \frac{1}{ \log \frac{r_A \kappa_A^2}{r_A \kappa_A^2-1} } \log \frac{1}{\epsilon} \right) 
\end{align}
It can be seen that if $r_A \kappa_A^2 \gg 1$, then the ratio $ \frac{r_A \kappa_A^2}{r_A \kappa_A^2-1} $ would approaches 1, so its log can be approximated as:
\begin{align}
    \log \frac{r_A \kappa_A^2}{r_A \kappa_A^2-1} = \log \left(1 + \frac{1}{r_A \kappa_A^2-1 } \right) \approx \frac{1}{r_A \kappa_A^2}
\end{align}
Therefore $T$ is upper bounded as $T = \mathcal{O}\left( r_A \kappa_A^2 \log \frac{1}{\epsilon}\right) $. We remark that this is an upper bound, and in reality, the value of $ \log \frac{r_A \kappa_A^2}{r_A \kappa_A^2-1} $ can be smaller, e.g., of $\mathcal{O}(1)$. In such a case, effectively, the value of $T$ is asymptotically of order $\mathcal{O}\left( \log \frac{1}{\epsilon} \right)$. 

At the same time, as $||A||_F^2 \geq r_A \sigma_{\min}^2$, following the same line of deduction as above, it can also be shown in an analogous manner that:
\begin{align}
     \frac{1}{ -\log \big(1 - \frac{\sigma^2_{\min}(A)}{||A||_F^2}  \big)} \geq \frac{1}{ \log \frac{r_A }{r_A -1} } \approx r_A
\end{align}
So, $T$ is lower bounded by $\Omega(r_A)$, and thus effectively, it holds that $T \in \mathcal{O}\left( r_A \log \frac{1}{\epsilon} \right)$.

\section{Complexity analysis}
\label{sec: analysis}
To analyze the complexity, we discuss the complexity step by step based on Algo.~\ref{algo: quantumkaczmarz}. Let $\mathcal{C}_k$ denote the circuit complexity of implementing the unitary $U_{x^{(k)}}$. Moreover, let $\mathcal{C}(a_{j_k})$ denote the circuit complexity of the unitary $U_{j_k}$ that prepares the row state $\ket{a_{j_k}}$ (with $j_k\in{1,2,...,n}$). With this notation, the cost of each step in Algo.~\ref{algo: quantumkaczmarz} is as follows:
\begin{itemize}
    \item \underline{Step $1$}: we need to use $U_{x^{(k)}},  U_{j_k}$ one time each. So, the total circuit complexity is $\mathcal{O}\left( \mathcal{C}_k + \mathcal{C}(a_{j_k}) \right)$. 
    \item \underline{Step $2$}: we use the $x$-rotation gate $R_x(\theta)$  one time, so the complexity is $\mathcal{O}(1)$. 
    \item \underline{Step $3$}: we use the unitary block-encodings from the previous two steps one time each, and another usage of the unitary $U_{j_k}$. So, the total complexity is:
    \begin{align}
        \mathcal{O}\left( \mathcal{C}_k + 2 \mathcal{C}(a_{j_k}) \right)
    \end{align}
    \item \underline{Step $4$}: we use the unitary block-encoding from the previous step, plus another use of the unitary $U_{x^{(k)}}$. The total circuit complexity is then:
    \begin{align}
        \mathcal{O}\left( 2\mathcal{C}_k + 2 \mathcal{C}(a_{j_k}) \right)
    \end{align}
    \item \underline{Step $5$}: this step uses the unitary block-encoding from the previous step $\mathcal{O}(1)$ (with Lemma \ref{lemma: amp_amp}), so the total circuit complexity is:
    \begin{align}
        \mathcal{O}\left( 2\mathcal{C}_k + 2 \mathcal{C}(a_{j_k}) \right)
    \end{align}
\end{itemize}
We remind that the outcome of the Step 5 (see Algo.~\ref{algo: quantumkaczmarz}) is the unitary block-encoding $U_{x^{(k+1)}}$ of an operator that has $x^{(k+1)}$ as the first column. The circuit  complexity of $ U_{x^{(k+1)}}$, as pointed out above, is:
\begin{align}
    \mathcal{C}_{k+1} = \mathcal{O}\left( 2\mathcal{C}_k + 2 \mathcal{C}(a_{j_k}) \right)
\end{align}
From here, we can use induction to proceed. Under similar reasoning, it can be shown that $\mathcal{C}_k = \mathcal{O}\left( 2\mathcal{C}_{k-1} + 2 \mathcal{C}(a_{j_{k-1}}) \right)$. For subsequent convenience, we define $\mathcal{C} \equiv  \max \{ \mathcal{C}(a_{j_k}) \}_{k=1}^T $. Then we have:
\begin{align}
\begin{split}
    \mathcal{C}_{k+1} &= \mathcal{O}\left( 2\mathcal{C}_k + 2 \mathcal{C}(a_{j_k}) \right) \\
    &= \mathcal{O}\left( 2(2\mathcal{C}_{k-1} + 2 \mathcal{C}(a_{{k-1}})) + 2 \mathcal{C}(a_{j_k}) \right) \\
    &= \mathcal{O}\left( 4\mathcal{C}_{k-1} + (2+4) \mathcal{C}\right) 
\end{split}
\end{align}
Similarly, $ \mathcal{C}_{k-1} = \mathcal{O}\left( 2\mathcal{C}_{k-2} + 2 \mathcal{C}(a_{j_{k-2}}) \right)$, so:
\begin{align}
\begin{split}
      \mathcal{C}_{k+1} &=  \mathcal{O}\left( 4\left( 2\mathcal{C}_{k-2} + 2 \mathcal{C}(a_{j_{k-2}}) \right) + (2+4) \mathcal{C} \right)  \\
      &= \mathcal{O}\left(  2^3 \mathcal{C}_{k-2} +  (2+2^2+2^3) \mathcal{C} \right)  \\
\end{split}
\end{align}
Proceeding further in a similar manner, it can be deduced that:
\begin{align}
\begin{split}
      \mathcal{C}_{k+1}
      &= \mathcal{O}\left(  2^{k+1} \mathcal{C}_{0} +  \sum_{i=1}^{k+1} 2^i \  \mathcal{C} \right) 
\end{split}
\end{align}
where $\mathcal{C}_0$ is the circuit complexity of $U_{x^{(0)}}$. For a total of $T$ iteration, we have the circuit complexity $\mathcal{C}_T$ of $U_{x^{(T)}}$ is:
\begin{align}
    \mathcal{O}\left(  2^{T} \mathcal{C}_{0} +  \sum_{i=1}^{T} 2^i \  \mathcal{C} \right) 
\end{align}
It is of well-known property that $ \sum_{i=1}^{T} 2^i  = 2^{T+1}-2 = \mathcal{O}(2^T)$, so the total complexity above is $ \mathcal{O}\left(  2^{T} \big( \mathcal{C}_{0} +  \mathcal{C} \big) \right) $. Finally, we measure the ancilla qubits on the state in Eqn.~\ref{measurement} and post-select on seeing $\ket{\bf 0}$. The success probability of this measurement is $||x_T||_2^2$, which can be quadratically improved via amplitude amplification method, incurring a further complexity $\mathcal{O}( \frac{1}{||x_T||_2})$. Accounting for this amplification and measurement step, we arrive at the totally final complexity $ \mathcal{O}\left( \frac{1}{||x_T||_2} 2^{T} \big( \mathcal{C}_{0} +  \mathcal{C} \big) \right) $. We remark that upon an appropriate choice of $T$, $x_T$ is a good approximation to the true solution $x_{\rm solution}$ of the linear system $Ax = b$. As analyzed in \cite{harrow2009quantum}, the inverse of the norm of $x_{\rm solution}$ is upper bounded as $\mathcal{O}(\kappa)$. Therefore, in the worst case, $\frac{1}{||x_T||_2} = \mathcal{O}(\kappa)$. 

Since $U_{x^{(0)}}$ can be arbitrary, it is safe to assume that its circuit complexity is $\mathcal{O}(1)$. Regarding $\mathcal{C}$, which is defined as $\max  \{ \mathcal{C}(a_{j_k}) \}_{k=1}^T$, according to Lemma \ref{lemma: statepreparation}, if for all $j$, the state $\ket{a_j}$ can be prepared via approaches in \cite{mcardle2022quantum, marin2023quantum, nakaji2022approximate, zoufal2019quantum}, the maximum circuit complexity (in terms of gate complexity) $ \mathcal{C} \in \mathcal{O}(\log m)$ (using constant number of ancilla qubits). In the same Lemma \ref{lemma: statepreparation}, it was stated that if instead $\ket{a_j}$ is prepared via \cite{zhang2022quantum} (which can work for arbitrary structure of $\ket{a_j}$), the depth complexity could be achieved as $\mathcal{O}\left( \max \{ \log s_{j_k} \}_{k=1}^T \right)$ (where $s_{j_k}$ is the number of nonzero entries of $a_{j_k}$, which is at most $m$) at the cost of employing extra $\mathcal{O}( \max \{ s_{j_k} \}_{k=1}^T ) $ ancilla qubits. So, we arrive at the final circuit complexity $\mathcal{O}\left( 2^T \log m\right)$ (using extra $\mathcal{O}(1)$ ancilla qubit). The circuit depth could be improved to $ \mathcal{O}\left( \max \{ \log s_{j_k} \}_{k=1}^T \right)$ using extra $\mathcal{O}( \max \{ s_{j_k} \}_{k=1}^T ) $ ancilla qubits. 

From the bounds of $T$ derived in the previous appendix, by replacing them to the final complexity, we arrive at the result stated in Theorem \ref{theorem: mainresult}.

\end{document}